\documentclass[11pt]{article}

\usepackage{owna4,amsmath,amstext,times}
\usepackage{graphicx}
\usepackage{algorithm}
\usepackage{algorithmic}

\renewcommand{\baselinestretch}{1.5}
\begin{document}

\noindent
\textbf{\LARGE SIFT: An Algorithm for Extracting Structural Information From Taxonomies}

\noindent
Jorge Martinez-Gil, \\
Software Competence Center Hagenberg (Austria), jorgemar@acm.org


\medskip

\section*{Abstract}
In this work we present SIFT, a 3-step algorithm for the analysis of the structural information represented by means of a taxonomy. The major advantage of this algorithm is the capability to leverage the information inherent to the hierarchical structures of taxonomies to infer correspondences which can allow to merge them in a later step. This method is particular relevant in scenarios where taxonomy alignment techniques exploiting textual information from taxonomy nodes cannot operate successfully. 

\section{Introduction}	

The problem of aligning taxonomies is one of the most interesting and relevant issues for knowledge engineers, since it has implications in a wide range of computational problems including file system merging, creation  of operating systems distributions, catalog integration, distributed taxonomy search, and so on. The non-deterministic nature of the problem is given by the fact that not even humans are able to identify optimal alignments \cite{DBLP:conf/icadl/Jung06a}, so the process is highly subjective. This means that its boundaries often go further than the category of an engineering problem what makes difficult to find closed solutions in this area. However, the large amount of people, research groups and resources dedicated to provide solutions in this field, tells us that we are facing a key challenge in order for a convincing way to automatically integrate taxonomic knowledge to become real.

In the last years, the need for methods to integrate knowledge has increased. Note that the need for aligning taxonomies comes from the old field of database schema integration \cite{DBLP:journals/vldb/RahmB01}. This field was born to work in a unified way with databases which had been developed independently. Nowadays researchers aim to make the techniques for aligning knowledge models flexible and powerful enough to work with all kind of database schemas, XML schemas, taxonomies, E/R models, dictionaries, and so on. Therefore, the problem we are facing consists of providing a set of correspondences between the nodes of two taxonomies about the same domain but which have been developed separately \cite{DBLP:journals/tkde/ShvaikoE13}.

The major contribution of this work is the proposal of a 3-step algorithm that is able to analyze the structural information represented by means of a taxonomy. The major advantage of this analysis is that it can allow us to leverage the information inherent to the hierarchical structure of taxonomies to infer correspondences which can allow to automatically merging them in a later step. This is particular relevant in scenarios where taxonomy matching techniques exploiting textual information from the taxonomy nodes cannot operate successfully.

From now on, this work is structured in the following way: The second section describes the state-of-the-art on taxonomy alignment. The third section describes the design and development of the algorithm. Case studies section provides some scenarios where our algorithm can help to solve real problems, including a brief discussion on the strengths and weaknesses of the proposal. Finally, we outline the key points of our contribution and propose future research tasks.

\section{Related Work}

The problem of aligning taxonomies have received much attention by the research community since various knowledge based applications, including clustering algorithms, browsing support interfaces, and recommendation systems, perform more effectively when they are supported with domain describing taxonomies, which help to resolve ambiguities and provide context \cite{DBLP:conf/odbis/CandanCSS08}. Furthermore, this problem is of great interest on a number of application areas, especially in scientific \cite{DBLP:conf/icadl/Jung06a}, business \cite{DBLP:conf/esws/AanenNVF12} \cite{DBLP:journals/pvldb/NandiB09}, and web data integration \cite{DBLP:journals/internet/GraciaM12} \cite{DBLP:conf/ijcai/PonzettoN09}. 

Taxonomy alignment techniques are able to detect taxonomy concepts that are equivalent. But, when can we say that two concepts are equivalent? If we attend only to the text label for representing the concepts, we can find many examples in everyday life, for instance, \textit{lift} and \textit{elevator} or \textit{car} and \textit{automobile} seem to be equivalent concepts since they represent the same real idea or object. However, it is well known that when taxonomies are used as knowledge sources, the way users perceive the degree of likeness between pairs of concepts is highly dependent on the domain being explored \cite{DBLP:conf/odbis/CandanCSS08}. Therefore, synonymy between text labels is not always an equivalence indicator, so it is necessary to focus in the context the concepts are being considered. 

Existing taxonomy alignment techniques focus on different dimensions of the problem, including whether data instances are used for matching \cite{DBLP:journals/sigmod/GilM10}, whether linguistic information and other auxiliary information are available \cite{DBLP:journals/isf/GilM13}, and whether the match is performed for complex structures \cite{DBLP:conf/icde/MelnikGR02}. Our algorithm fits in this last category.

Algorithms implementing techniques for matching complex structures are mostly based on heuristics. Heuristics consider, for example, that elements of two distinct taxonomies are similar if their direct sub-concepts, and/or their direct super-concepts and/or their brother concepts are similar \cite{DBLP:conf/semweb/ReynaudS06}. These structural techniques can be based on a fixed point like that proposed in \cite{DBLP:conf/vldb/MadhavanBR01}, or can be viewed as a satisfiability problem of a set of propositional formulas \cite{DBLP:conf/semweb/AvesaniGY05}. There are also some proposals to align taxonomies supposed to be asymmetric from a structural point of view \cite{DBLP:conf/semweb/HamdiSNR10}, or to create matching functions by means of a composition of various techniques designed to make best use of the characteristics of the taxonomies \cite{DBLP:conf/semweb/ReynaudS06}.

Despite such advances in matching technologies, taxonomy alignments using linguistic information and other auxiliary information are rarely perfect \cite{DBLP:journals/air/Gil14}. In particular, imperfection can be due to homonyms (i.e., nodes with identical concept-names, but possibly different semantics) and synonyms (concepts with different names but same semantics). However, the major advantage of pure structural matching techniques is that finding perfect alignments is possible in many cases.

\section{Contribution}
We approach the problem from the classic perspective, that it is to say, a taxonomy can be defined as a set of concepts that have been hierarchically organized to control the terms belonging to a vocabulary. The goal is to facilitate a number of operations on items from a repository. However, a problem occurs when two item repositories have to be merged, since it is also necessary to merge the two taxonomies which describe them. 

Our contribution to face this problem is the proposal of an efficient 3-step algorithm for the analysis of taxonomies describing such repositories. This analysis could be helpful for solving the problem of heterogeneity between the given taxonomies from a strictly structural point of view in a later step. As a collateral effect, the output data from our algorithm could be also used for exploiting any kind of solution involving the use of information from the structure of the given taxonomies. Use cases section will explore this in more detail.

More formally, we can define a mapping as an expression that can be written in the form \textit{(c, c', n, R)}. Where \textit{c} and \textit{c'} are concepts belonging to different taxonomies, \textit{R} is the relation of correspondence and \textit{n} is a real number between 0 and 1. \textit{n} represents the degree of confidence for \textit{R}. In our work, \textit{c} and \textit{c'} will be concepts represented by means of taxonomy nodes (a.k.a. taxons) which are assigned a rank and can be placed at a particular level in a systematic hierarchy reflecting relationships. Moreover, the relation \textit{R} which describe how \textit{c} and \textit{c'} are related is going to be of similarity. 

The algorithm that we propose is divided into three high level steps. The first step is optional since it is only necessary when the given knowledge model is not a taxonomy yet, but another kind of more general model like an graph or an ontology \cite{DBLP:journals/ijseke/SunLL12}.

\begin{enumerate}
	\item To convert the knowledge model into a taxonomy (See Algorithm 1).
	\item	To store the taxonomy in some parts of a special data structure (See Algorithm 2).
	\item	To order and fill the data structure with complementary calculations (See Algorithm 3).
\end{enumerate}

Finally, it is necessary to call the algorithm (See Algorithm 4).
The philosophy of the algorithm consists of detecting the changes in the depths of each taxon in the hierarchy. In this way, it is possible to count the different kinds of neighbors that a concept may have.

Before designing the algorithm, it is also necessary to define a data structure (DS) to store the data calculated by the algorithm.
The data structure is a linked list with six records in each node: depth, children, brothers, brothers\_left, same\_level and name. Table 1 tells us the data type and a brief description of each of these records. In the next subsections, we are going to describe more in depth each of the main steps of the proposed algorithm.

\begin{table}
\centering
    \begin{tabular}{|l|l|l|}
        \hline
        \textbf{Attribute}    & \textbf{Type }   & \textbf{Description  }     \\ \hline
        depth        & integer & Level of the current taxon (begins with 0)   \\ 
        children     & integer & Number of children of the current taxon      \\ 
        brothers     & integer & Number of brothers of the current taxon      \\ 
        brothersLeft & integer & Number of brother taxons that are above this \\ 
        sameLevel    & integer & Number of taxons with the same depth         \\ 
        name         & string  & ID of the taxon                              \\
        \hline
    \end{tabular}
		\caption{A node of the linked list which stores the information}
\end{table}

\subsection{Converting a knowledge model into a taxonomy}
This is the first step which consists of converting the model into a taxonomy which will allow us to compute more easily the data related to the neighborhood of each concept into the knowledge model. This step is optional and it is only necessary when the input is not a perfect hierarchy but contains some cycles. This is the usual case when working with graph models or ontologies, for example. The procedure is inspired by one proposed in \cite{DBLP:journals/internet/McBride02} to visit all the concepts in an ontology. Algorithm 1 shows the related portion of pseudocode.

\renewcommand{\baselinestretch}{1}

\begin{algorithm}
\begin{algorithmic}[1]
\REQUIRE cls: \textit{class}, occurs: \textit{list}, depth: \textit{integer}
\STATE storingInTax(cls, depth); \textbf{Step 2}
\IF {(cls.\textit{canAs}( \textit{model}.class ) \textbf{AND} (\textbf{NOT} occurs.\textit{contains}( cls )))}
\WHILE {iterator = cls.SubClasses}
\STATE \textit{class} sub := (\textit{class}) iterator.next
\STATE occurs.\textit{add}(cls)
\STATE \textit{ont2tax} (sub, occurs, depth + 1)
\STATE occurs.\textit{remove}(cls)
\ENDWHILE
\ENDIF
\RETURN \TRUE
\end{algorithmic}
\caption{ont2tax: Procedure for converting a generic knowledge model into a taxonomy}\label{alg:algo1}
\end{algorithm}

\renewcommand{\baselinestretch}{1.5}

\subsection{Storing the taxonomy in the data structure}
In this second step, we only know the depth (number of indents for the taxon) and the name of each concept, so we can only partially fill the data structure, thus, we can only invoke the procedure with the arguments depth and concept name.

\renewcommand{\baselinestretch}{1}

\begin{algorithm}
\begin{algorithmic}[1]
\REQUIRE cls: \textit{ontology}, depth: \textit{integer}
\STATE \textit{Element} e := new \textit{Element} (depth, 0, 0, 0, 0, cls.getName)
\STATE DS.\textit{add} (e)
\RETURN \TRUE
\end{algorithmic}
\caption{storingInTax: Storing the taxonomy in the data structure}\label{alg:algo2}
\end{algorithm}

\renewcommand{\baselinestretch}{1.5}

\subsection{Ordering and filling the data structure}
With data stored in the DS, we can now detect the changes in the depth of the entries in the taxonomy to compute the number of children, brothers and, so on.
It is necessary to take into account the following rules:

\begin{enumerate}
	\item All taxons with the same depth are same level taxons.
	\item	A chain of brothers is a chain of taxons at the same level.
	\item	A change to an outer taxon breaks a chain of brothers.
	\item	All brothers with a previous position are on the left.
	\item	Given a taxon, if the following concept has an inner depth, it is a child.
	\item	A chain of children can only be broken by a change to an outer taxon.
	\item	An inner taxon (grandson taxon) does not break a chain of children.
\end{enumerate}

Algorithm 3 shows us the procedural implementation for this set of rules. The computational complexity of this procedure is low, even in the worst of cases we would have $O(n^{2})$, since the most complex portion of code can be implemented by means of two simple loops. This means that our solution presents a great scalability regardless of the platform on which the algorithm could be implemented and executed.

\renewcommand{\baselinestretch}{1}

\begin{algorithm}
\begin{algorithmic}[1]
\REQUIRE children, brothers, brothers left: integer
\REQUIRE same level, i, j, k, t: integer
\REQUIRE ag: boolean
\FOR{i := 0 to DS.size}
   \STATE children, brothers, brothers left := 0

   \FOR{j := 0 to DS.size}
    \IF{if (j $<$ i)}
       \IF{if (DS[i].depth = DS[j].depth)}
         \STATE brothers++
         \STATE brothers left++
       \ENDIF
      \IF{(DS[i].depth $<$ DS[j].depth)}
         \STATE brothers := 0
         \STATE brothers left := 0
      \ENDIF
   \ENDIF
   \IF{(j $>$ i)}
      \IF{(DS[i].depth = DS[j].depth)}
         \STATE brothers++
      \ENDIF
      \IF{(DS[i].depth $<$ DS[j].depth)}
         \STATE break
      \ENDIF
   \ENDIF
   \IF{ ((j = i+1)  AND (DS[i].depth = DS[j].depth - 1) AND (NOT  ag))}
      \FOR {for k := j to DS[j].depth $<$ DS[k].depth}
         \IF{(DS[j].depth = DS[k].depth)}
            \STATE child++
            \STATE ag := true
         \ENDIF
      \ENDFOR
   \ENDIF
   \ENDFOR
   \FOR{for t := 0 to DS.size}
      \IF{if (NOT t=i) AND (DS[i].depth = DS[t].depth)}
         \STATE same level++
      \ENDIF
   \ENDFOR
   \STATE DS[i].addNumChildren (children)
   \STATE DS[i].addNumBrothers (brothers)
   \STATE DS[i].addNumBrothersOnTheLeft (brother left)
   \STATE DS[i].addNumSameLevel (same level)
\ENDFOR
\RETURN \TRUE
\end{algorithmic}
\caption{finalStep: Ordering and filling the data structure}\label{alg:algo3}
\end{algorithm}

\renewcommand{\baselinestretch}{1.5}

\subsection{Calling to the algorithm}
Now, it is necessary to invoke the algorithm. At this point it is necessary to define the taxonomy model and to locate the concepts without ancestors, in order to begin to visit all the concepts. This is particular relevant in forest models\footnote{Forest model is that kind of graph model where there is no connection between some graph components}. Note that the ArrayList is necessary to store the visited concepts. Algorithm 4 shows the related portion of pseudocode.

\renewcommand{\baselinestretch}{1}

\begin{algorithm}
\begin{algorithmic}[1]
\STATE Model m := \textit{createModel}
\STATE Iterator i := m.\textit{listHierarchyRootClasses}()
\WHILE {i.\textit{hasNext}()}
   \STATE onto2tax((Class) i.\textit{next}(), new ArrayList(), 0)
\ENDWHILE
\STATE \textit{finalStep} ()
\end{algorithmic}
\caption{calling to the 3-step algorithm}\label{alg:algo4}
\end{algorithm}

\renewcommand{\baselinestretch}{1.5}

\section{Case studies}

The purpose of this section is to show the relative ease with which a taxonomy analysis can be performed or a new taxonomy matcher can be developed, based on the data obtained from the algorithm. In the next subsections we are going to show three use cases: how to use the algorithm to compute the leaves in a taxonomy, how to use it to obtain the structural index of a taxonomy, and finally how to use it to align taxonomies automatically.

\subsection{Computing the number of leaves in a taxonomy}

There are techniques that compute the leaves in a graph for performing a graph analysis. In this sense, our algorithm is easy to extend in order to compute the number of leaves in a taxonomy. To do so, it is only necessary to compute the number of the deepest taxons. We are going to see how to compute the leaves of the taxonomy for an example but, it is possible to compute other features such as paths. Algorithm 5 shows us how to compute the leaves (i.e. terminal nodes) of a given taxonomy.

\renewcommand{\baselinestretch}{1}

\begin{algorithm}
\begin{algorithmic}[1]
\REQUIRE var max, leaves: integer
\STATE max := leaves := 0
\FOR { i := 0 to DS.size}
   \IF {(DS[i].depth $>$ max)}
      \STATE max := DS[i].depth
   \ENDIF
\ENDFOR
\FOR {for j := 0 to DS.size}
   \IF{(DS[j].depth = max)}
      \STATE leaves++
   \ENDIF
\ENDFOR
\RETURN leaves
\end{algorithmic}
\caption{leaves: computing the leaves of a taxonomy}\label{alg:algo5}
\end{algorithm}

\renewcommand{\baselinestretch}{1.5}

\subsection{Comparing structural similarities} 

It is possible to use our algorithm for extracting structural indexes of taxonomies in order to compare its structural similarity. The structural index of a taxonomy is a kind of hash function that tells global information about the total number of children, brothers and so on.

As we show in the state-of-the-art, some techniques use statistical methods for obtaining the structural similarity. It can be useful for adjusting the quality of the generated mappings, for example.

Algorithm 6 shows how to automatically compute one possible structural index from a taxonomy.

\renewcommand{\baselinestretch}{1}

\begin{algorithm}
\begin{algorithmic}[1]
\REQUIRE var acum : integer
\STATE acum := 0
\FOR {i := 0 to DS.size}
   \STATE acum := acum + DS[i].depth
   \STATE acum := acum + DS[i].children
   \STATE acum := acum + DS[i].brothers
   \STATE acum := acum + DS[i].leftbrothers
   \STATE acum := acum + DS[i].samelevel
\ENDFOR
\RETURN acum
\end{algorithmic}
\caption{structuralIndex: extract a structural index of the ontology}\label{alg:algo6}
\end{algorithm}

\renewcommand{\baselinestretch}{1.5}

Obviously, when comparing two structural indexes, the higher percentage, the higher the structural similarity of the compared taxonomies. This means that if two taxonomies share the same structural index, we can state that its structural organization is equivalent.

\subsection{Real alignment situations}
Our algorithm also allows that information to be obtained from the analysis phase can be helpful in order to take decisions in taxonomy alignment scenarios. Output data from SIFT allow us to easily create customized rule-based matchers to obtain more accurate taxonomy alignments. For example, the similarity between two taxonomy concepts or taxons could be given by certain rules concerning ancestors, brothers, and so on.

Moreover, it is possible to combine our proposal with other basic matching algorithms. This can be done by designing a formula that may allow us to align taxonomies from the point of view of the elements, and from the taxonomy structure. This is possible due to the fact one of the attributes (name) contains information at the element level, so it is possible to exploit this kind of information by using some kind of computational method like the Levenshtein algorithm \cite{levenshtein1966} which is able to calculate similarity between two text strings. In this way, if many attributes (whether structural or textual) are similar, the concepts are also supposed to be similar.

\section{Conclusions \& Future Work}

In this work, we have designed and implemented, SIFT that is a 3-step algorithm that allows us to analyze the structural information inherent to the hierarchical structures of taxonomies. This can be useful when solving problems concerning heterogeneity between taxonomies describing a same domain but which have been developed separately. Therefore, the algorithm that we propose is valid for taxonomy alignment, but also for aligning ontologies, directory listings, file systems, operating system distributions,  and in general whatever kind of model which can be transformed into a taxonomy. Our algorithm tries to leverage the inherent characteristics from taxonomies to infer correspondences which can allow us to merge them in a later step, even without text labels describing each of the nodes from the taxonomy.

As future work, we should work to leverage the good performance of our algorithm by designing a combined alignment strategy. In this work, we have proposed to use each of the attributes with similar weights. However, this strategy could not be optimal in some specific cases. We aim to redefine this strategy so that a preliminary study should try to automatically determine the kind of problem we are facing at a given moment, and dynamically assign higher weights to the most promising taxon attributes. 

\section*{Source Code}
An implementation of this algorithm can be found at \url{https://github.com/jorgemartinezgil/sift}

\bibliographystyle{abbrv}
\bibliography{bib}

\end{document}